\documentclass{PoS}

\usepackage{amsmath,amssymb}

\newcommand{\mpios}{M_\pi^\mathrm{OS}}

\title{Isospin-0 $\pi\pi$ scattering from twisted mass lattice QCD}

\ShortTitle{Isospin-0 $\pi\pi$ scattering}

\author{\speaker{L.~Liu} $^1$, S.~Bacchio $^{2,3}$, P.~Dimopoulos $^{4,5}$ , J.~Finkenrath $^{6}$, R.~Frezzotti $^{7}$, C.~Helmes $^1$, C.~Jost $^1$, B.~Knippschild $^1$, B.~Kostrzewa $^1$, H.~Liu $^8$, K.~Ottnad $^1$,  M.~Petschlies $^1$,  C.~Urbach $^1$,M.~Werner $^1$\\
 \llap{$^1$}Helmholz-Institut f\"ur Strahlen- und Kernphysik and Bethe Center for Theoretical
             Physics, Universit\"at Bonn, D-53115 Bonn, Germany\\
\llap{$^2$}Department of Physics, University of Cyprus, PO Box 20537, 1678 Nicosia, Cyprus\\
\llap{$^3$}Fakult\"at f\"ur Mathematik und Naturwissenschaften, Bergische Universit\"at Wuppertal, 42119 Wuppertal, Germany \\
\llap{$^4$}Centro Fermi - Museo Storico della Fisica e Centro Studi e Ricerche
Enrico Fermi, Compendio del Viminale, Piazza del
Viminiale 1, I-00184, Rome, Italy \\
\llap{$^5$}Dipartimento di Fisica, Universit{\`a} di Roma ``Tor Vergata",
Via della Ricerca Scientifica 1, I-00133 Rome, Italy \\
\llap{$^6$}Computation-based Science and Technology Research Center, The Cyprus Institute, PO Box 27456, 1645 Nicosia, Cyprus \\
\llap{$^7$}Dipartimento di Fisica, Universit{\`a} and INFN di Roma Tor Vergata, 00133 Roma, Italy\\
\llap{$^8$} Albert Einstein Center for Fundamental Physics, University of Bern, 3012 Bern, Switzerland

 E-mail: \email{liuming@hiskp.uni-bonn.de}}

\abstract{We present results for the isospin-0 $\pi\pi$ s-wave scattering length calculated in twisted mass lattice
QCD. We use three $N_f = 2$ ensembles with unitary pion mass at its physical value, 240~MeV and 330~MeV respectively. We also use a large set of $N_f = 2 + 1 +1$ ensembles with unitary pion masses varying in the range
of 230~MeV - 510~MeV at three different values of the lattice
spacing. A mixed action approach with the Osterwalder-Seiler action in the valence sector is adopted to circumvent the complications arising from isospin symmetry breaking of the twisted mass quark action. Due to the relatively large lattice artefacts in the $N_f = 2 + 1 +1$ ensembles, we do  not present the scattering lengths for these ensembles. Instead, taking the advantage of the many different pion masses of these ensembles, we qualitatively discuss the pion mass dependence of the scattering properties of this channel based on the results from the $N_f = 2 + 1 +1$ ensembles. The scattering length is computed for the $N_f = 2$ ensembles and the chiral extrapolation is performed. At the physical pion mass, our result $M_\pi a^\mathrm{I=0}_0 = 0.198(9)(6)$ agrees reasonably well with various experimental measurements and theoretical predictions. }

\FullConference{34th annual International Symposium on Lattice Field Theory\\
		24-30 July 2016\\
		University of Southampton, UK}

\begin{document}

\section{Introduction}
Elastic $\pi\pi$ scattering is a fundamental QCD process at low energies. It provides an ideal testing ground for the mechanism of chiral symmetry breaking. The isospin-0  $\pi\pi$ scattering is particularly interesting because it accommodates
the lowest resonance in QCD -- the mysterious $\sigma$ or $f_0(500)$ scalar meson. Studying this channel in lattice QCD is difficult mainly due to the fermionic disconnected diagrams contributing to the isospin-0 $\pi\pi$ correlation function. To date there are only two full lattice QCD computations dedicated to this channel ~\cite{Fu:2013ffa, Briceno:2016mjc}. In this work we compute the scattering length of the isospin-0 $\pi\pi$ channel in twisted mass lattice QCD~\cite{Frezzotti:2000nk}. We use a mixed action approach with the Osterwalder-Seiler (OS) action~\cite{Frezzotti:2004wz} in the valence sector to circumvent the complications arising from the isospin symmetry breaking of the twisted mass quark action. 


\section{Lattice setup}
\label{sec:actions}
The results presented in this paper are based on the gauge
configurations generated by the European Twisted Mass Collaboration
(ETMC).
We use three $N_f = 2$ ensembles with Wilson clover twisted mass quark 
action at maximal twist~\cite{Frezzotti:2000nk}. The pion masses of the three ensembles are at the physical value, 240~MeV and 330~MeV,
respectively. The lattice spacing is $a=0.0931(2)\ \mathrm{fm}$ for
all three ensembles. More details about these emsembles are presented in Ref.~\cite{Abdel-Rehim:2015pwa}.
In addition, we use a set of $N_f = 2
+ 1 + 1$ ensembles with Wilson twisted mass quark action with pion masses varying in the range
of 230~MeV - 510~MeV at three different values of the lattice
spacing~\cite{Baron:2010bv,Baron:2010th}.  We follow the notation in
these references and denote the ensembles as A, B, and D ensembles
with lattice spacing values $a_A = 0.0863(4)\ \mathrm{fm}$, 
$a_B = 0.0779(4)\ \mathrm{fm}$ and $a_D = 0.0607(2)\ \mathrm{fm}$,
respectively.  

In Table~\ref{tab:setup} we list all ensembles used in this study with the relevant input
parameters, the lattice volume and the number of configurations.

\begin{table}[t!]
 \centering
 \begin{tabular*}{.9\textwidth}{@{\extracolsep{\fill}}lccccccc}
  \hline\hline
  ensemble & $\beta$ & $c_{\mathrm{sw}}$ &$a\mu_\ell$ & $a\mu_\sigma$ & $a\mu_\delta$ &
  $(L/a)^3\times T/a$ & $N_\mathrm{conf}$  \\ 
  \hline\hline
  $cA2.09.48$ &2.10 &1.57551 &0.009 &- &- &$48^3\times96$ & $615$ \\
  $cA2.30.48$ &2.10 &1.57551 &0.030  &- &- &$48^3\times96$ & $352$ \\
  $cA2.60.32$ &2.10 &1.57551 &0.060  &- &- &$32^3\times64$ & $337$ \\
  \hline
  $A30.32$   & $1.90$ &- & $0.0030$ & $0.150$  & $0.190$  &
  $32^3\times64$ & $274$  \\
  $A40.24$   & $1.90$ &- & $0.0040$ & $0.150$  & $0.190$  &
  $24^3\times48$ & $1017$  \\
  $A40.32$   & $1.90$ &- & $0.0040$ & $0.150$  & $0.190$  &
  $32^3\times64$ & $251$  \\
  $A60.24$   & $1.90$  &- & $0.0060$ & $0.150$  & $0.190$  &
  $24^3\times48$ & $314$  \\
  $A80.24$   & $1.90$ &- & $0.0080$ & $0.150$  & $0.190$  &
  $24^3\times48$ & $307$  \\
  $A100.24$  & $1.90$ &- & $0.0100$ & $0.150$  & $0.190$  &
  $24^3\times48$ & $313$  \\
  \hline
  $B25.32$   & $1.95$ &- & $0.0025$ & $0.135$  & $0.170$  &
  $32^3\times64$ & $201$ \\
  $B55.32$   & $1.95$ &- & $0.0055$ & $0.135$  & $0.170$  &
  $32^3\times64$ & $311$ \\
  $B85.24$   & $1.95$ &- & $0.0085$ & $0.135$  & $0.170$  &
  $32^3\times64$ & $296$ \\
  \hline
  $D15.48$ & $2.10$ &- & $0.0015$ & $0.120$ & $0.1385$ &
  $48^3\times96$ & $313$ \\
  $D30.48$ & $2.10$ &- & $0.0030$ & $0.120$ & $0.1385$ &
  $48^3\times96$ & $198$ \\
  $D45.32sc$ & $2.10$ &- & $0.0045$ & $0.0937$ & $0.1077$ &
  $32^3\times64$ & $301$ \\

  \hline\hline
 \end{tabular*}
 \caption{The gauge ensembles used in this study. The labelling of
   the ensembles follows the notations in
   Ref.~\cite{Abdel-Rehim:2015pwa, Baron:2010bv}. In addition to the relevant input
   parameters we give the lattice volume $(L/a)^3\times T/a$ and  the number of evaluated
   configurations $N_\mathrm{conf}$.}
 \label{tab:setup}
\end{table}

In the valence sector we introduce quarks in the so-called
Osterwalder-Seiler (OS) discretisation~\cite{Frezzotti:2004wz}. 
The OS  \textit{up} and \textit{down} quarks have explicit SU$(2)$ isospin symmetry 
if the proper parameters of the actions are chosen. 
The matching of OS to unitary actions is performed by matching the quark
mass values. Masses computed with OS valence quarks differ from those computed with
twisted mass valence quarks by lattice artefact of
$\mathcal{O}(a^2)$, in particular
$(M_\pi^\mathrm{OS})^2 - (M_\pi)^2\ =\ \mathcal{O}(a^2)$. 
For twisted clover fermions this difference is much reduced as
compared to twisted mass fermions~\cite{Abdel-Rehim:2015pwa}, however,
the effect is still sizable. We use the OS pion mass in this paper,
with the consequence that the pion mass values of all ensembles are higher than the values 
measured in the unitary theory.  

As a smearing scheme we use the stochastic Laplacian Heavyside (sLapH)
method~\cite{Peardon:2009gh,Morningstar:2011ka} for our
computation. The details of the sLapH parameter choices for the
$N_f = 2+1+1$ Wilson twisted mass ensembles are given in
Ref.~\cite{Helmes:2015gla}. The parameters for the $N_f = 2$ ensembles are the same as those for $N_f = 2+1+1$ ensembles with the
corresponding lattice volume.

\section{L\"uscher's finite volume method}
\label{sec:FiniteVolumeMethod}
L\"uscher
showed that the infinite volume scattering parameters can be
related to the discrete spectrum of the eigenstates in a finite-volume 
box~\cite{Luscher:1986pf, Luscher:1990ux}. In the case of s-wave elastic scattering,
L\"uscher's formula reads:
 $ q \cot \delta_0 (k) = \mathcal{Z}_{00}(1; q^2)/\pi^{3/2}\,$,
where $k$ is the scattering momentum and $q$ is a dimensionless variable defined via 
$q= k L/ 2\pi$.  $\mathcal{Z}_{00}(1; q^2)$ is the L\"uscher zeta-function which
can be evaluated numerically given the value of $q^2$. Using the
effective range expansion of s-wave elastic scattering near 
threshold, we have 
 $ k \cot \delta_0 (k) = \frac{1}{a_0} + \frac{1}{2} r_0 k^2 + \mathcal{O}(k^4)$,
where $a_0$ is the scattering length and $r_0$ is the effective
range parameter. Once the isospin-0 $\pi\pi$ interacting energy $E_{\pi\pi}$ is determined from
lattice QCD simulations, the scattering length $a_0$ can be calculated
from the following relation
\begin{equation}
  \label{eq:ScatLen}
  \frac{2\pi}{L} \frac{\mathcal{Z}_{00}(1; q^2)}{\pi^{3/2}} =
  \frac{1}{a_0} + \frac{1}{2} r_0 k^2 + \mathcal{O}(k^4)\,.
\end{equation}

\section{Finite volume spectrum}
\label{sec:FiniteVolumeSpectrum}
The discrete spectra of hadronic states are extracted
from the correlation functions of the interpolating operators that
resemble the states. We define the interpolating operator that represents the isospin-0
$\pi\pi$ state in terms of OS valence quarks
\begin{equation}
  \label{Eq:pipioperator}
  \mathcal{O}_{\pi\pi}^\mathrm{I=0}(t) = \frac{1}{\sqrt{3}}(
  \pi^{+}\pi^{-}(t)\ +\ \pi^{-}\pi^{+}(t)\ +\
  \pi^{0}\pi^{0}(t)), 
\end{equation}
with single pion operators summed over spatial coordinates
$\mathbf{x}$ to project to zero momentum 
  \begin{equation}
   \label{eq:piop}
 \pi^{+}(t) = \sum_\mathbf{x} \bar{d} \gamma_5 u
    (\mathbf{x},t), 
    \pi^{-}(t) = \sum_\mathbf{x} \bar{u} \gamma_5 d (\mathbf{x},t) ,
    \pi^{0}(t) = \sum_\mathbf{x} \frac{1}{\sqrt{2}} (
    \bar{u}\gamma_5 u - \bar{d}\gamma_5 d)(\mathbf{x},t).
  \end{equation}
Here $u$ and $d$ represent the OS {\it up} and {\it down} quarks,
respectively. With OS valence quarks all three pions are mass
degenerate and will be denoted as $\mpios$. 

\begin{figure}[t]
 \centering
 \begin{tabular}{cccc}
  \includegraphics[width=0.15\linewidth]{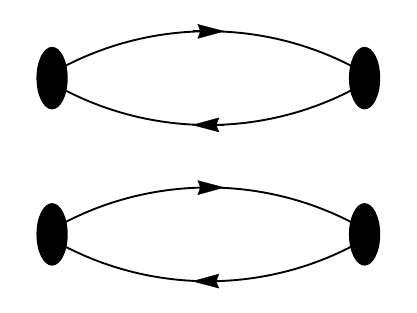}&\includegraphics[width=0.15\linewidth]{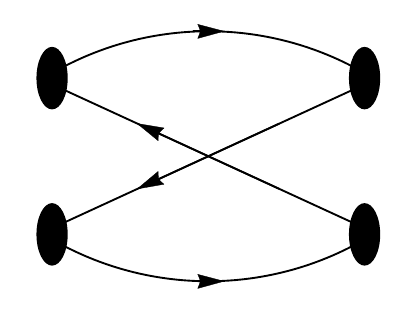}&\includegraphics[width=0.15\linewidth]{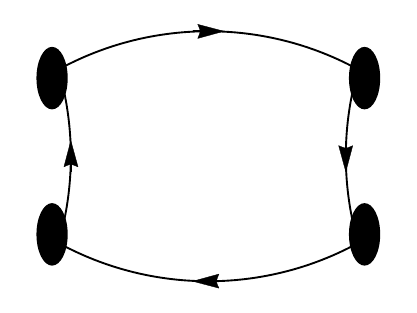}&\includegraphics[width=0.15\linewidth]{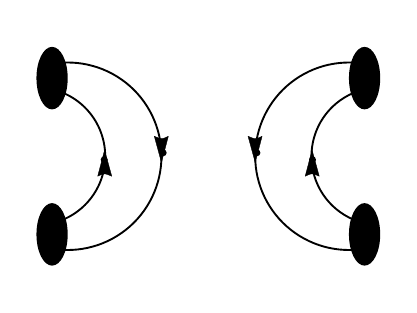} \\
  $D(t)$ &$X(t)$ &$B(t)$ &$V(t)$ 
    \end{tabular}
  \caption{Diagrams contributing to the correlation
    function $C_{\pi\pi}(t)$.}
  \label{fig:diagrams}
\end{figure}

\begin{figure}[t]
  \centering
  \includegraphics[width=0.45\linewidth]{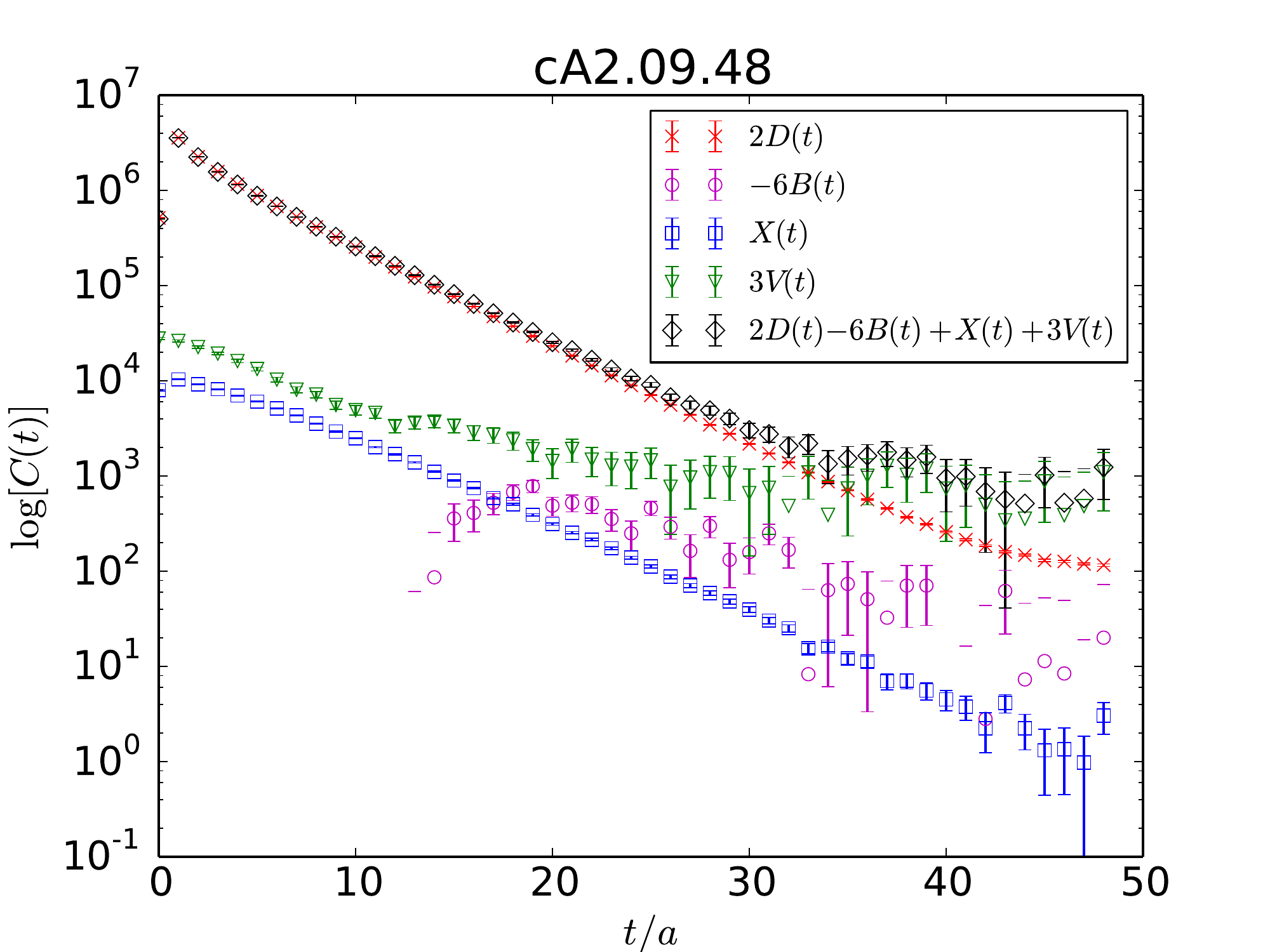}\includegraphics[width=0.45\linewidth]{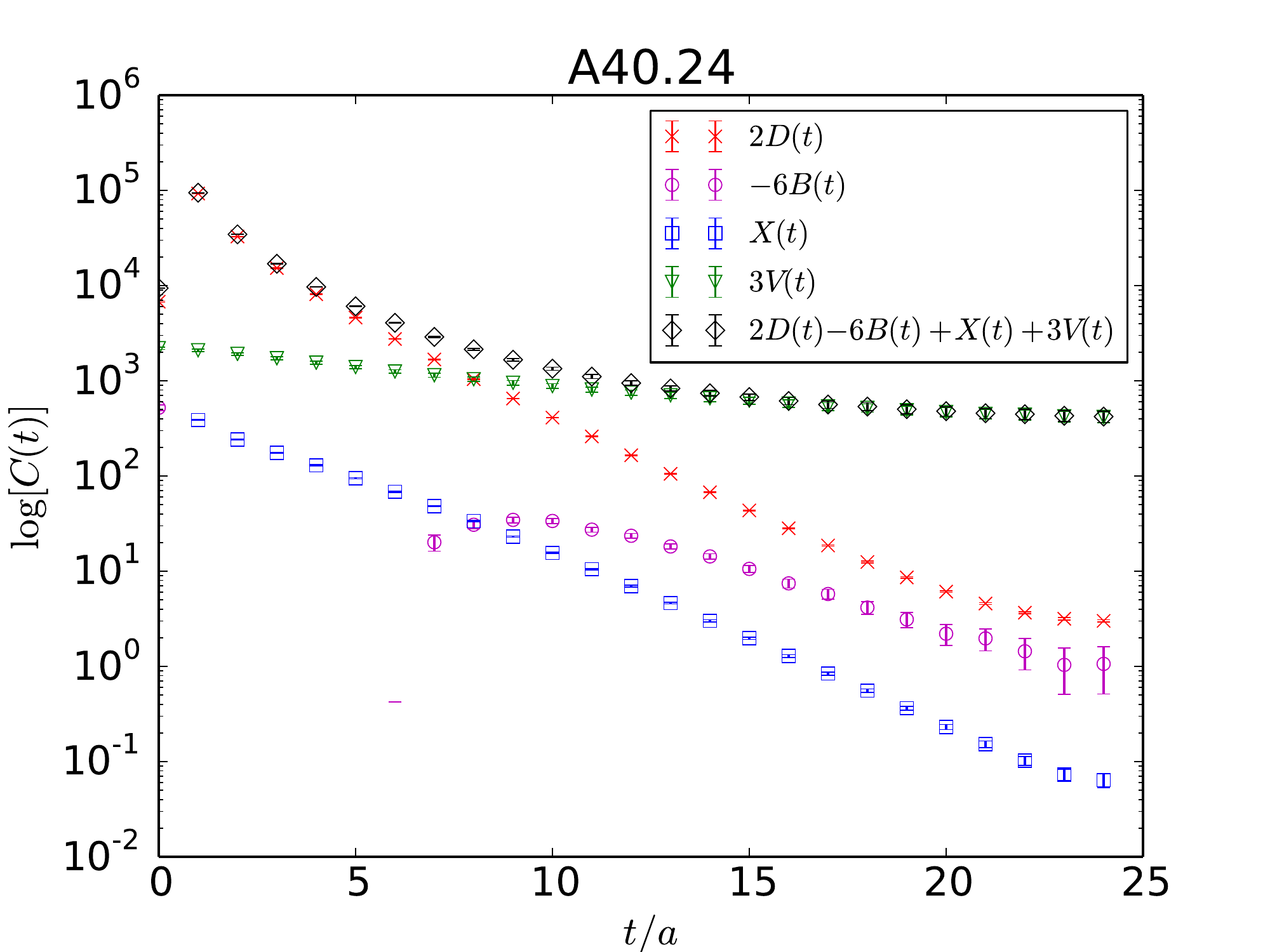} 
  \caption{Correlation functions of the operator
    $\mathcal{O}_{\pi\pi}^\mathrm{I=0}$ and the single diagrams
    $D,X,B,V$ for the ensembles $cA2.09.48$ and $A40.24$.}
  \label{fig:corr}
\end{figure}

The energy of the isospin-0 $\pi\pi$ state can be computed from the
exponential decay in time of the correlation function
$  C_{\pi\pi}(t) = \frac{1}{T}\sum_{t_{src}=0}^{T-1}\langle
  \mathcal{O}_{\pi\pi}^\mathrm{I=0} (t + t_{src})\  ( 
  \mathcal{O}_{\pi\pi}^\mathrm{I=0} )^\dagger (t_{src}) \rangle\,$,
where $T$ is the temporal lattice extend. The four diagrams
contributing to this correlation function, namely the direct connected
diagram $D(t)$, the cross diagram $X(t)$, the box diagram $B(t)$ and
the vacuum diagram $V(t)$, are depicted in
Fig.~\ref{fig:diagrams}. The correlation function can be 
expressed in terms of all relevant diagrams as 
 $C_{\pi\pi}(t) = 2D(t) + X(t) - 6B(t) + 3V(t)$.
$C_{\pi\pi}$ and the contributions from individual diagrams $D,X,B$
and $V$ are plotted in Fig.~\ref{fig:corr} for the 
ensembles $cA2.09.48$ and $A40.24$ as examples. 

Even though we have full SU$(2)$ isospin symmetry in the valence
sector when using OS valence quarks as described above, we have to
consider effects of unitarity breaking. In particular, the $n (n \geq 1)$ unitary neutral pion
states can mix with the operator $ \mathcal{O}_{\pi\pi}^\mathrm{I=0}$ via the 
vacuum diagram $V$. 
Since the neutral pion is the lightest meson in the spectrum
with Wilson twisted mass fermions at finite lattice spacing,
the appearance of such states with $n=1$ (and maybe $n=2$) will dominate
the large Euclidean time behaviour of the correlation function
$C_{\pi\pi}$. The effect of this mixing can be clearly seen in the plot of the correlation function
$C_{\pi\pi}(t)$ for the ensemble $A40.24$ (the right panel of Fig.~\ref{fig:corr}), in which the vacuum diagram 
starts to dominate $C_{\pi\pi}(t)$ at around $t=10$. While for the ensemble $cA2.09.48$, this effect is much less
prominent. Please note that this mixing is purely a lattice artefact. Since the lattice artefacts of the twisted clover action
are much smaller than of the twisted mass action~\cite{Abdel-Rehim:2015pwa}, we expect that the effect of the unitary 
neutral pion mixing is much smaller for the $N_f = 2$ ensembles compared to the $N_f = 2 + 1 + 1$ ensembles. 
This is indeed what we observed generally for the $N_f = 2$ ensembles and the $N_f = 2 + 1 + 1$ ensembles used in this work.
In order to resolve this mixing, we build
a $2\times 2$ matrix of correlation functions 
$  C_{ij}(t) = \frac{1}{T}\sum_{t_{src}=0}^{T-1} \langle
  \mathcal{O}_{i}(t + t_{src})
  \mathcal{O}_{j}^\dagger (t_{src}) \rangle$, 
with $i,j$ labeling the operator $\mathcal{O}_{\pi\pi}^\mathrm{I=0}$
and the unitary neutral pion operator 
$  \pi^{0, uni}(t) = \sum_\mathbf{x} \frac{1}{\sqrt{2}}( \bar{u}\gamma_5
  u\ - \  \bar{d^\prime}\gamma_5 d^\prime)(\mathbf{x},t)\, $, 
where $u$ and $d^\prime$ are the (unitary) Wilson (clover) twisted mass
{\it up} and {\it down} quarks. We use $d^\prime$ to distinguish it
from OS {\it down} quark in Eq.~\ref{eq:piop}. The twisted
mass {\it up} quark coincides with the OS {\it up} quark with our
matching scheme of the OS to the unitary action.  

We use a shifting procedure $\tilde{C}_{ij}(t) =
C_{ij}(t) - C_{ij}(t+1)$ 
to eliminate contaminations constant in time from so-called 
thermal states due to the finite time extension of the 
lattice.  
Solving the generalized eigenvalue problem(GEVP) 
 $ \tilde{C}(t)\, v(t,t_0) = \lambda(t,t_0)\, \tilde{C}(t_0)\, v(t,t_0)\,,$
the desired energy of the $\pi\pi$ isospin-0 system $E_{\pi\pi}$ can
be extracted from the exponential decay of the eigenvalues
$\lambda(t,t_0)$. To further improve our results we adopt a method to remove excited state
contaminations, which we have recently used
successfully to study $\eta$ and $\eta^\prime$
mesons~\cite{Jansen:2008wv,Michael:2013gka}. See Ref.~\cite{Liu:2016cba} for more details
about this method. 
In Table~\ref{tab:EnergyResults}, we collect
the values of $E_{\pi\pi}$ obtained from the procedure described above. The OS pion masses $\mpios$ are also
given since they will be needed to compute the scattering length. 


\begin{table}[t!]
  \centering
  \begin{tabular*}{\textwidth}{@{\extracolsep{\fill}}lcc|cccc}
    \hline\hline
    Ensemble   & $a \mpios$ & $a E_{\pi\pi}$  &Ensemble   & $a \mpios$ & $a E_{\pi\pi}$ \\
    \hline
    $cA2.09.48$ &0.11985(15) &0.2356(4) & $B85.24$  &0.2434(6) &0.4441(29)  \\
    $cA2.30.48$ &0.15214(11) &0.3010(3) &$A30.32$ &0.2143(10) &0.4043(36) \\
    $cA2.60.32$  &0.18844(24) &0.3647(5) &$A40.24$ &0.2283(10) &0.4187(40)  \\
    $D15.48$ &0.1082(3) &0.2067(13)  &$A40.32$ &0.2266(8) &0.4417(20) \\
    $D30.48$ &0.1299(2) &0.2526(8)  &$A60.24$ &0.2482(9) &0.4644(39)  \\
    $D45.32$ &0.1466(6) &0.2686(17)   &$A80.24$ &0.2663(7) &0.5033(21)  \\
    $B25.32$ &0.1843(13) &0.3496(31)   &$A100.24$ &0.2835(6) &0.5096(55)  \\
    $B55.32$ &0.2105(4) &0.4051(23)   \\
    \hline\hline
  \end{tabular*}
  \caption{OS pion masses and the $\pi\pi$ interacting energies in lattice units for the
    three ensembles.}
  \label{tab:EnergyResults}
\end{table}

\section{Results}
\label{sec:results}

\begin{figure}[t]
  \includegraphics[width=0.5\linewidth]{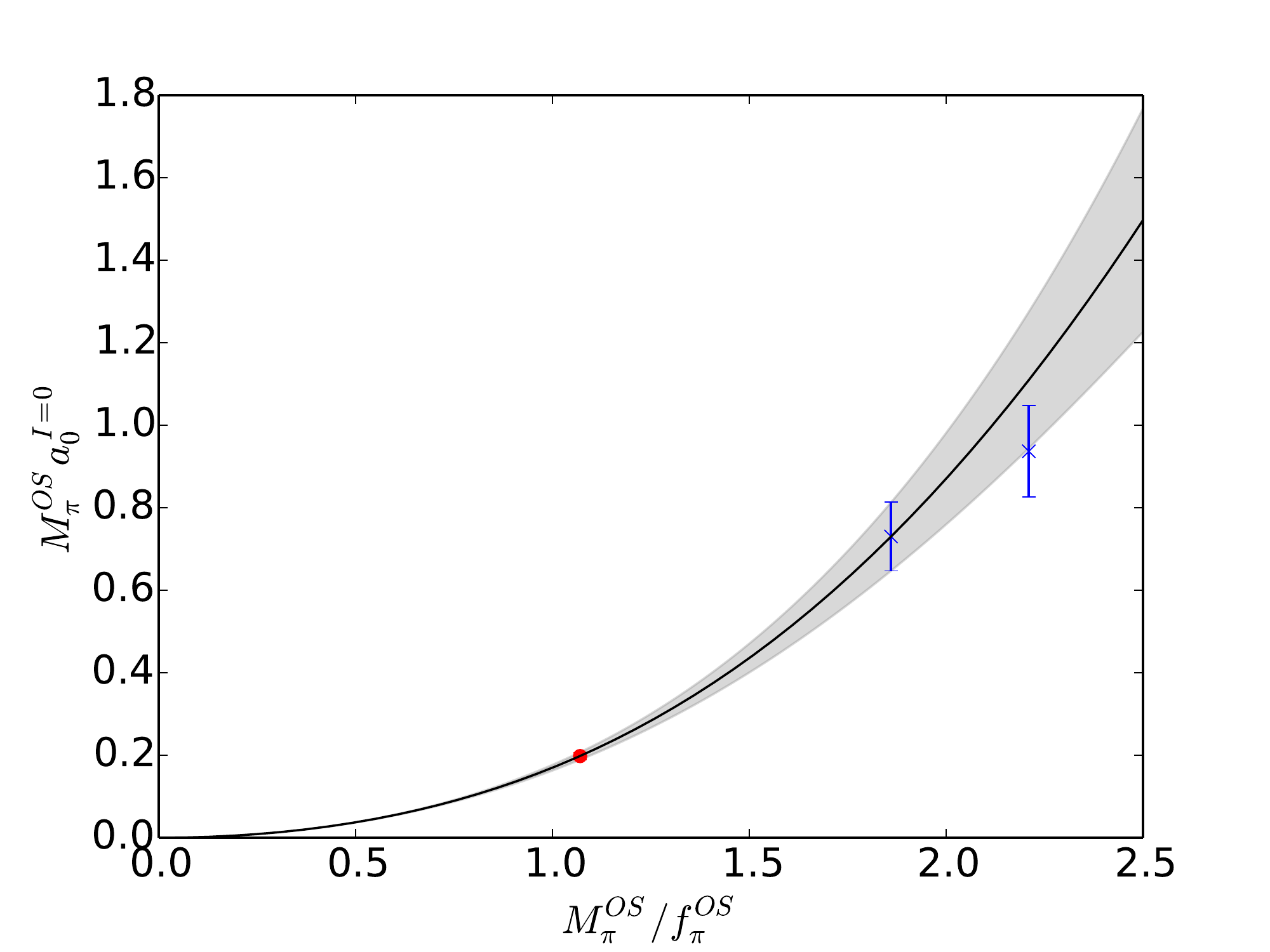}\includegraphics[width=0.5\linewidth]{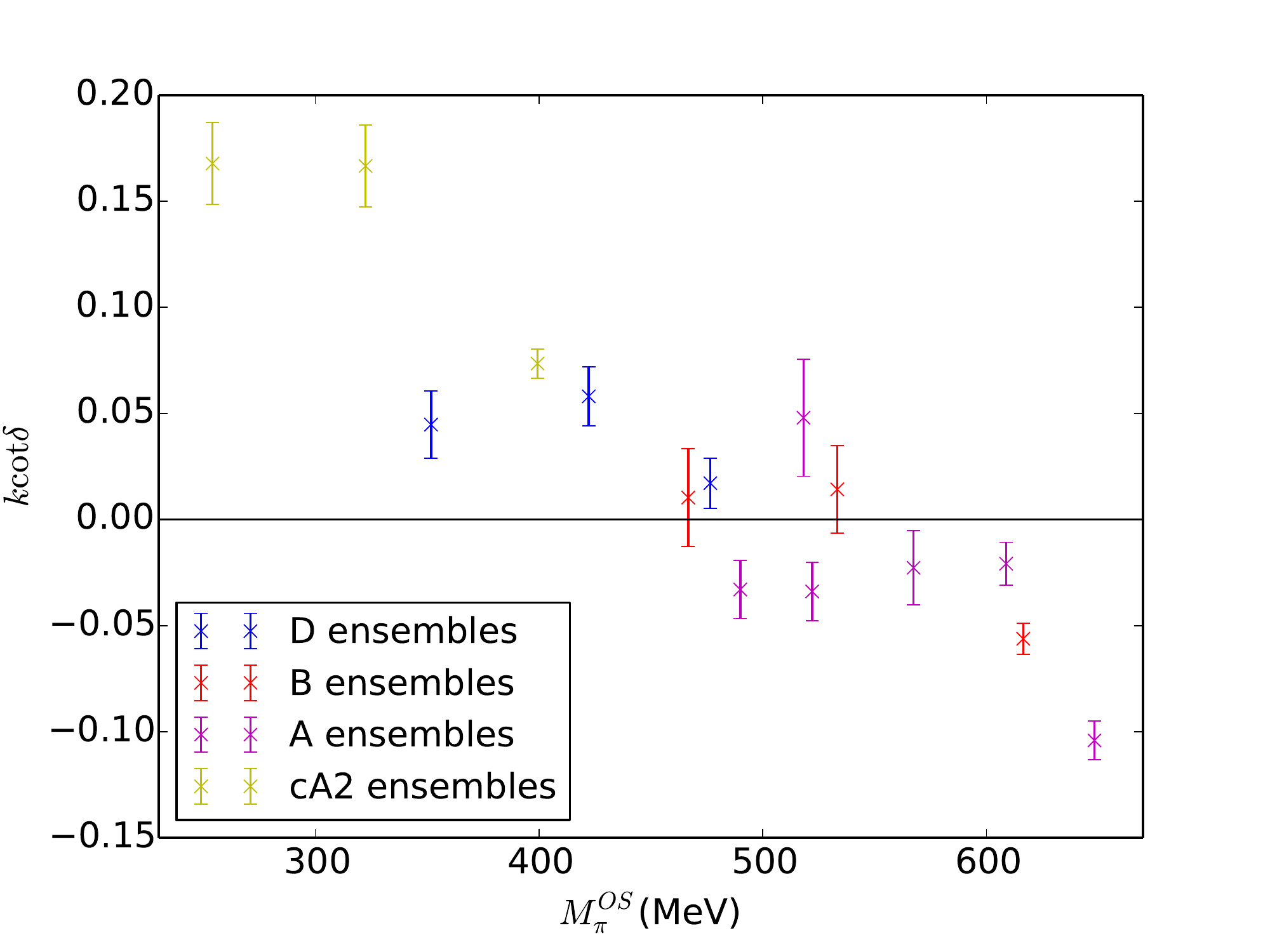} 
  \caption{(a): The values of $\mpios a_0^\mathrm{I=0}$ for the ensembles $cA2.09.48$ and $cA2.60.32$. The 
  black curve and the grey band represent the chiral fit using only the data point with lower pion mass.
  The red point indicates the extrapolated
  value at physical pion mass. (b): The values of $k\cot\delta(k)$ for all ensembles as a function of $\mpios$. }
  \label{fig:results}
\end{figure}

The scattering momentum $k^2$ is calculated from
 the energies $E_{\pi\pi}$ and the OS
pion masses listed in Table~\ref{tab:EnergyResults}. Then the
scattering length can be obtained from
Eq.~\ref{eq:ScatLen}. Using the values of the effective range $r_0$ determined
from $\chi$PT~\cite{Gasser:1983yg, Liu:2016cba}, we investigated 
the contribution of $\mathcal{O}(k^2)$ term in the effective range expansion.
For the ensembles $cA2.09.48$ and $cA2.30.48$, the value of $\frac{1}{2}r_0 k^2$
is less than $3\%$ of $k\cot\delta(k)$. So we can safely ignore the $\mathcal{O}(k^4)$ term
and compute the scattering length $a_0^\mathrm{I=0}$ using Eq.~\ref{eq:ScatLen} for these two ensembles. 
The values of  $\mpios a_0^\mathrm{I=0}$ are plotted in Fig.~\ref{fig:results}(a) as a function of 
$\mpios/f_{\pi}^\mathrm{OS}$. For the ensemble $cA2.60.32$, the contribution of $\frac{1}{2}r_0 k^2$
is rather large -- around $30\%$ of $k\cot\delta(k)$. Since the contribution of $\mathcal{O}(k^4)$ is unclear,
we refrain from giving the scattering length for this ensemble. 
The reason for the invalidity of the effective range
expansion is probably due to virtual or bound state poles appearing in the isospin-0
$\pi\pi$ scattering amplitude at the pion mass around 400~MeV,
which is the OS pion mass of the ensemble cA2.60.32. Since the OS pion mass for the 
$N_f = 2 + 1 + 1$ ensembles are generally above 400~MeV, we do not compute the 
scattering length for these ensembles either. However, the value of $k\cot\delta(k)$ can 
be computed up to lattice artefacts. Fig~\ref{fig:results}(b) presents the values of  $k\cot\delta(k)$ for all ensembles as a function of $\mpios$. One can see that $k\cot\delta(k)$ changes from positive to negative with increasing OS pion mass.  
The pion mass range where the sign change happens is around 400~MeV - 600~MeV. Correspondingly, the scattering length will change from positive infinity to negative infinity in this range, which indicates the emergence of virtual or bound state poles in the scattering amplitude.

For the $N_f = 2$ ensembles, chiral extrapolation is performed in order to obtain the scattering length at the physical pion mass. Since we only have
two data points, we fit the NLO $\chi$PT formula, which contains one free
parameter, to our data. 
The method we are applying here is valid only in the elastic
region. Therefore, the pion mass values must be small enough to be
below threshold where the $\sigma$ meson becomes stable. 
Furthermore, the pion mass value should also be small enough to make the 
chiral expansion valid. To be safe, we perform the chiral extrapolation 
using only the data point with the lower pion mass (~250~MeV). The fit 
results using the two data points are used to estimate the systematics 
arising from chiral extrapolation.
This leads to our 
final result for the scattering length:
\begin{equation}
M_\pi a^\mathrm{I=0}_0 = 0.198(9)_\mathrm{stat}(6)_\mathrm{sys}\,.
\end{equation}

\section{Summary and discussions}
The isospin-0 $\pi\pi$ scattering is studied with
L\"uscher's finite volume formalism in twisted mass lattice  
QCD using a mixed action approach with the OS action in the valence sector. 
The lowest energy level in the rest frame is extracted for
three $N_f =2$ ensembles and a large set of
$N_f=2+1+1$ ensembles with many different values of pion mass. 
The scattering length is computed for the two $N_f=2$ ensembles with the lowest pion mass values. 
After the chiral extrapolation, our result at the physical pion mass is $M_\pi a^{I=0}_0 = 0.198(9)(6)$ , which is
compatible with the newer experimental and theoretical
determinations available in the literature. The value of $k\cot\delta(k)$ near threshold is computed for all ensembles.
The pion mass dependence of the scattering properties of this channel is briefly discussed.  We cannot exclude that our
result is affected by residual systematic
uncertainties stemming from unitarity breaking, which will vanish in
the continuum limit. In order
to avoid isospin breaking and unitarity breaking effects, we will
repeat this computation with an action without isospin breaking.

\begin{acknowledgments}
We thank the
members of ETMC for the most enjoyable collaboration. The computer
time for this project was made available to us by the John von
Neumann-Institute for Computing (NIC) on the Jureca and Juqueen
systems in J{\"u}lich. 
We thank A.~Rusetsky and Zhi-Hui Guo for very useful discussions and R.~Brice\~no for
useful comments.
This project was funded by the DFG as a project in
the Sino-German CRC110. S.~B. has received funding from the Horizon 2020 research and innovation 
program of the European Commission under the Marie Sklodowska-Curie
programme GrantNo. 642069. This work was granted access to the HPC resources IDRIS under the
allocation 52271 made by GENCI.
The open source software
packages tmLQCD~\cite{Jansen:2009xp}, Lemon~\cite{Deuzeman:2011wz}, DD$\alpha$AMG~\cite{Alexandrou:2016izb}
and
R~\cite{R:2005} have been used.
\end{acknowledgments}

\bibliographystyle{h-physrev4.bst}
\bibliography{bibliography}

\begin{thebibliography}{10}

\bibitem{Fu:2013ffa}
Z.~Fu,
\newblock Phys. Rev. {\bf D87}, 074501 (2013), [1303.0517].

\bibitem{Briceno:2016mjc}
R.~A. Briceno, J.~J. Dudek, R.~G. Edwards and D.~J. Wilson,
\newblock 1607.05900.

\bibitem{Frezzotti:2000nk}
ALPHA, R.~Frezzotti, P.~A. Grassi, S.~Sint and P.~Weisz,
\newblock JHEP {\bf 08}, 058 (2001), [hep-lat/0101001].

\bibitem{Frezzotti:2004wz}
R.~Frezzotti and G.~C. Rossi,
\newblock JHEP {\bf 10}, 070 (2004), [hep-lat/0407002].

\bibitem{Abdel-Rehim:2015pwa}
ETM, A.~Abdel-Rehim {\em et~al.},
\newblock 1507.05068.

\bibitem{Baron:2010bv}
ETM, R.~Baron {\em et~al.},
\newblock JHEP {\bf 06}, 111 (2010), [1004.5284].

\bibitem{Baron:2010th}
ETM, R.~Baron {\em et~al.},
\newblock Comput.Phys.Commun. {\bf 182}, 299 (2011), [1005.2042].

\bibitem{Peardon:2009gh}
Hadron Spectrum, M.~Peardon {\em et~al.},
\newblock Phys. Rev. {\bf D80}, 054506 (2009), [0905.2160].

\bibitem{Morningstar:2011ka}
C.~Morningstar {\em et~al.},
\newblock Phys. Rev. {\bf D83}, 114505 (2011), [1104.3870].

\bibitem{Helmes:2015gla}
ETM, C.~Helmes {\em et~al.},
\newblock JHEP {\bf 09}, 109 (2015), [1506.00408].

\bibitem{Luscher:1986pf}
M.~L{\"u}scher,
\newblock Commun.Math.Phys. {\bf 105}, 153 (1986).

\bibitem{Luscher:1990ux}
M.~L{\"u}scher,
\newblock Nucl.Phys. {\bf B354}, 531 (1991).

\bibitem{Jansen:2008wv}
ETM, K.~Jansen, C.~Michael and C.~Urbach,
\newblock Eur.Phys.J. {\bf C58}, 261 (2008), [0804.3871].

\bibitem{Michael:2013gka}
ETM, C.~Michael, K.~Ottnad and C.~Urbach,
\newblock Phys.Rev.Lett. {\bf 111}, 181602 (2013), [1310.1207].

\bibitem{Liu:2016cba}
L.~Liu {\em et~al.},
\newblock 1612.02061.

\bibitem{Gasser:1983yg}
J.~Gasser and H.~Leutwyler,
\newblock Ann. Phys. {\bf 158}, 142 (1984).

\bibitem{Jansen:2009xp}
K.~Jansen and C.~Urbach,
\newblock Comput.Phys.Commun. {\bf 180}, 2717 (2009), [0905.3331].

\bibitem{Deuzeman:2011wz}
ETM, A.~Deuzeman, S.~Reker and C.~Urbach,
\newblock 1106.4177.

\bibitem{Alexandrou:2016izb}
C.~Alexandrou {\em et~al.},
\newblock accepted by PRD  (2016), [1610.02370].

\bibitem{R:2005}
{R Development Core Team},
\newblock {\em R: A language and environment for statistical computing},
\newblock R Foundation for Statistical Computing, Vienna, Austria, 2005,
\newblock {ISBN} 3-900051-07-0.

\end{thebibliography}


\end{document}